\documentclass[fleqn,10pt]{wlscirep}
\usepackage[utf8]{inputenc}
\usepackage[T1]{fontenc}
\usepackage[export]{adjustbox}
\title{Gold-polymer hybrid metasurface for polarization-independent enhanced third harmonic generation in the ultraviolet}

\author[1,*]{Shroddha Mukhopadhyay}
\author[2]{Ana Conde-Rubio}
\author[1]{Jose Trull}
\author[2]{Agustin Mihi}
\author[3]{Michael Scalora}
\author[4]{Maria Antonietta Vincenti}
\author[1]{Crina Cojocaru}

\affil[1]{Department of Physics, Universitat Politècnica de Catalunya, Rambla Sant Nebridi 22,
08222 Terrassa (Barcelona), Spain}
\affil[2]{Institute of Materials Science of Barcelona ICMAB – CSIC,
08193 Bellaterra (Barcelona), Spain}
\affil[3]{Aviation and Missile Center, US Army CCDC, Redstone Arsenal, AL 35898-5000,
USA}
\affil[4]{Department of Information Engineering – University of Brescia, Via Branze 38,
25123 Brescia, Italy}

\affil[*]{shroddha.mukhopadhyay@upc.edu}

\begin{abstract}
We present a combined experimental and theoretical study of nonlinear light-matter interactions in a three-dimensional gold-polymer hybrid metasurface. In contrast to conventional two-dimensional designs, which by symmetry may support either transverse electric (TE) or transverse magnetic (TM) polarization, our volumetric architecture accepts both TE and TM modes simultaneously, reflecting the dimensionality and versatility required by the photonic devices. The metasurface
comprises a periodic lattice of gold nanostructures embedded in a dielectric polymer matrix, creating complex metal–dielectric interfaces that sustain tightly confined plasmonic resonances. When driven by ultrafast near-infrared pulses, these resonances concentrate optical energy at the nanoscale, enabling efficient third-harmonic generation and upconversion of visible light into the ultraviolet (UV) and deep-UV regimes with enhanced conversion efficiency. We perform spatial and temporal mapping of the nonlinear response under both TE and TM excitation. Our measurements reveal polarization-agnostic field enhancement and spectral tunability arising from the three-dimensional morphology—capabilities unattainable in planar metasurfaces, where two-dimensional symmetry inherently limits polarization flexibility and functional bandwidth. This 3D platform provides a flexible design toolbox for polarization-independent UV and deep-UV light sources. Potential applications include high-resolution UV spectroscopy, optical multiplexing, data storage, and emerging quantum photonic architectures. By establishing fundamental insights into three-dimensional nonlinear metasurface behavior, our work paves the way for next-generation reconfigurable, multi-polarization nanophotonic devices.
\end{abstract}
\begin{document}

\flushbottom
\maketitle
%
%
\thispagestyle{empty}

\section*{Introduction}

Nonlinear frequency conversion in metals has been a cornerstone of nonlinear optics research since its inception,
propelled by the large third-order susceptibilities of plasmonic materials despite their inherent absorption in the visible (VIS)
and near-infrared (NIR) bands \cite{bib1,bib2,bib3,bib4}. Historically, investigations have centered on planar or two-dimensional (2D)
metastructures, like gratings and pillar arrays, whose symmetry confines device functionality to either TE or TM
polarization. While these 2D platforms have yielded valuable insights, they fall short of representing the volumetric nature
of practical photonic components used in practical photonic systems.
Advances in nanofabrication now make possible to produce three-dimensional (3D) metal-dielectric architectures that
transcend the planar limit. By embedding metallic nanolayers within dielectric matrices and introducing depth-dependent
patterning, these metasurfaces localize electromagnetic fields in all three spatial dimensions. This field confinement not only
supports both TE and TM modes simultaneously but also enables richer light-matter interactions under ultrafast excitation,
unlocking higher-order nonlinear processes with unprecedented spatial and temporal control. 3D metastructures offer a
versatile design toolbox for next-generation photonic devices. Key advantages include: polarization-independent operation,
accepting and enhancing both TE and TM inputs without additional optical components; enhanced spectral tunability
through depth-engineered resonances, allowing dynamic wavelength conversion from visible to deep-UV regimes; improved
integration with volumetric platforms such as optical fibers, waveguide circuits, and compact sensors, where planar films
alone cannot deliver full functionality.
Despite rapid progress in fabricating 3D nanostructures and metamaterials, theoretical models have yet to fully capture
their volumetric nonlinear response under high-intensity, ultrashort pulses. Prior studies on second- and third-harmonic
generation from ultrathin gold layers (20–70 nm) illuminated intrinsic material\cite{bib3} but were intrinsically limited by
subwavelength interaction lengths with very low NL conversion efficiencies from simple metallic nanolayers, and planar geometries.

Nevertheless, corrugated nanostructures featuring surface reliefs on the order of 10 to 100 nm are known to modify the spatial distribution of an incident optical electromagnetic (EM) field. Such structures can be engineered to enhance light–matter interactions on purpose-built metasurfaces by exploiting plasmonic resonances. Surface plasmons (SP) are collective oscillations of free charges on conductive surfaces that couple strongly with an incident EM wave when excited at their resonant wavelength \cite{bib5,bib6,bib7}. Under resonant conditions, plasmonic structures can tightly confine the EM field near the surface of the metal, with the confinement determined by geometrical factors (such as shape, size, and sharpness) as well as physical properties (dielectric functions at a given wavelength) of both the metal and its surrounding dielectric materials. This strong plasmonic coupling significantly boosts light–matter interactions and generates large surface nonlinearities \cite{bib7,bib8,bib9,bib10,bib11,bib12,bib13}, enabling higher harmonic generation (HHG) under relatively low input laser intensities and paving the way for manipulating optical fields with unprecedented spatial and temporal resolution.

Based on the highly precise techniques now achievable in nano-fabrication, plasmonic structures can be built entirely from metal or as metal–dielectric composites. Much of the literature has focused on metallic nanostructures that simultaneously offer an intrinsic nonlinear optical (NLO) response and amplify the incident EM field. Examples include arrays for SHG using bow-tie configurations \cite{bib14}, V-shaped nanoantennas coupled with nanorod systems \cite{bib15}, L-shaped nanoparticles \cite{bib16}, split-ring resonators \cite{bib17}, rectangular hole arrays \cite{bib18}, and pyramidal recesses \cite{bib19}, among others. However, metal-dielectric composite structures allow us to take advantage of the complementary linear and NL optical properties of each material, combining the sharply localized SP resonances of metal nanolayers with the inherent strong NLO response of certain dielectrics. Consequently, a growing effort is dedicated to hybrid structures that combine a NL material (serving as the harmonic source) with a plasmonic nanostructure (acting as the harmonic amplifier). Traditionally, hybrid nanostructures have been predominantly applied to boost SHG, since purely metallic plasmonic nanoantennas lack measurable second-order nonlinearity. In contrast, plasmonic structures are notably more efficient in facilitating $\chi^{(3)}$ processes. By analyzing the electric field distribution around these nanostructures, one gains insight into the nature of the excited plasmons and the spectral selectivity of the field enhancement. Moreover, high-harmonic generation from hybrid structures can broaden the range of applications, from compact nanoscale extreme UV sources for microscopy, lithography, and spectroscopy to even attosecond physics, thereby opening up a myriad of new opportunities\cite{bib9,bib10,bib11,bib12,bib13}. For instance, Chen et al.\cite{bib10} introduced a periodic two-dimensional array of gold discs embedded in an organic conjugated polymer, demonstrating a spectrally selective THG with up to three times enhancement with respect to that of a flat polymer on gold. Similarly, Ren et al.\cite{bib11} examined a 100 nm thick gold nano-patch array coated with a polycarbonate polymer film doped with a polymethine dye, reporting a maximum THG enhancement factor of about 20 relative to that of a polymer film over a flat gold film. In another study, Albrecht et al. \cite{bib12} showcased a simpler plasmonic system wherein arrays of non-interacting, rod-shaped gold nanoparticles on a polymethyl methacrylate (PMMA) film exhibited conversion efficiency enhancements that tracked the spectral profile of the plasmonic resonance. Also, in an Au nano-rod/Cu$_2$O core/shell hybrid nanostructure, Bar-Elli et al. showed a THG enhancement of one order of magnitude, over bare Au nano-rod cores\cite{bib13}.

In this work, we investigate a gold-polymer hybrid metasurface featuring periodic patterning in the lateral plane and
depth-dependent morphology along the vertical axis. The structure combines a nanohole array atop a gold nanodisk lattice
within a polymer host, creating a quasi-3D crystal that supports efficient third-harmonic generation (THG) into the
UV under both TE and TM excitation. By mapping the spatial and temporal nonlinear response with ultrafast near-
infrared pulses, we demonstrate polarization-independent frequency upconversion far exceeding the performance of flat gold
films.

Although the crystal is not uniform along the z direction, it does not exhibit complete periodicity either, rendering it a “quasi-3D” system. Previous studies on similar quasi-3D plasmonic structures have highlighted their sensitivity, efficiency, and biocompatibility (attributed to gold) as refractometric sensors \cite{bib20,bib21,bib22,bib23}, and the standard nanofabrication technique used to fabricate the hybrid metasurface (details in section\ref{sec:3}) is quite well established. In our work, we exploit the narrow spectral linewidth resonance of the crystal to achieve a significant enhancement in the NL surface properties of gold at the nanoscale. This enhanced nonlinearity is experimentally quantified in terms of the THG from the patterned crystal, compared to that of an unpatterned gold nanolayer, demonstrating that the structure functions as an efficient light source in the deep UV range. The experimental observations are further corroborated and elucidated by theoretical calculations and simulations.

\section{\label{sec:2} Gold–polymer hybrid structure design}

\begin{figure*}
    \includegraphics[width=0.9\textwidth,center]{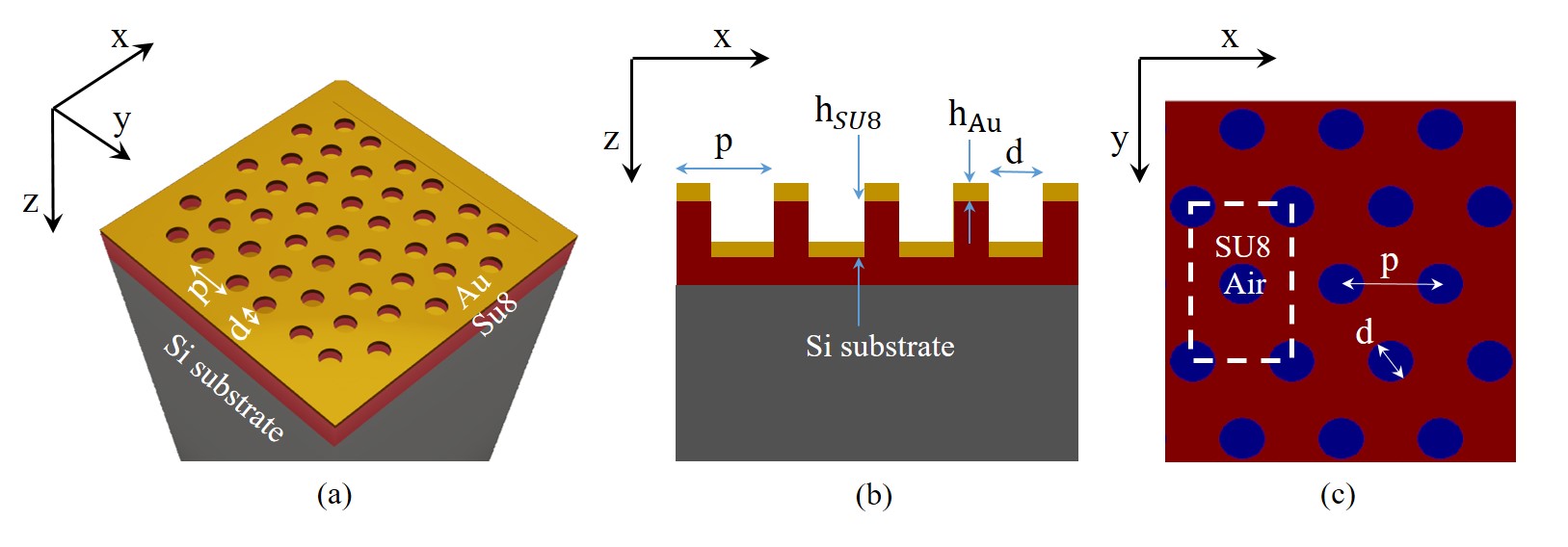}
\caption{\label{fig1} Sample schematic representation (a) 3D view of the quasi-3D hybrid structure, with the following dimensions $p=800 nm, h_{Au}=50 nm, d=367 nm$, and $h_{SU8}=390 nm$ (b) zoom in side view displaying the two layers of periodicity i.e. the upper layer of thin gold film with a nanohole array and the underlying gold nanodisk array, (c) top view scheme of the hexagonal lattice, with the dashed rectangle showing one unit cell of the hexagonal lattice, as modeled in the computation. }
\end{figure*}

The metasurface is composed of large-area, periodic arrays of cylindrical wells imprinted into an SU8 photoresist layer of a thickness of $\sim600 nm$, deposited on top of a silicon substrate. Each well has a diameter of $d=367 nm$ and a depth of $h_{SU8}=390nm$, with a center-to-center spacing between the holes of $p=800 nm$, and the pattern is periodic along the x and y directions, arranged in a hexagonal lattice. A uniform gold layer (with thickness $h_{Au} = 50 nm$) covers the entire nanostructure - both the top surface and the recessed regions - resulting in a quasi-3D plasmonic crystal. As shown in Fig.\ref{fig1}(a) and Fig.\ref{fig1}(b), the top layer consists of a 2D array of nano-holes in the gold film, while a second, distinct level of isolated gold disks is formed at the bottom of the embossed wells. Fig.\ref{fig1}(b) presents a side view, and Fig.\ref{fig1}(c) displays a top view of the hexagonal lattice, where the dashed rectangle shows one unit cell of the hexagonal lattice.

\begin{figure*}
    \includegraphics[width=\textwidth,center]{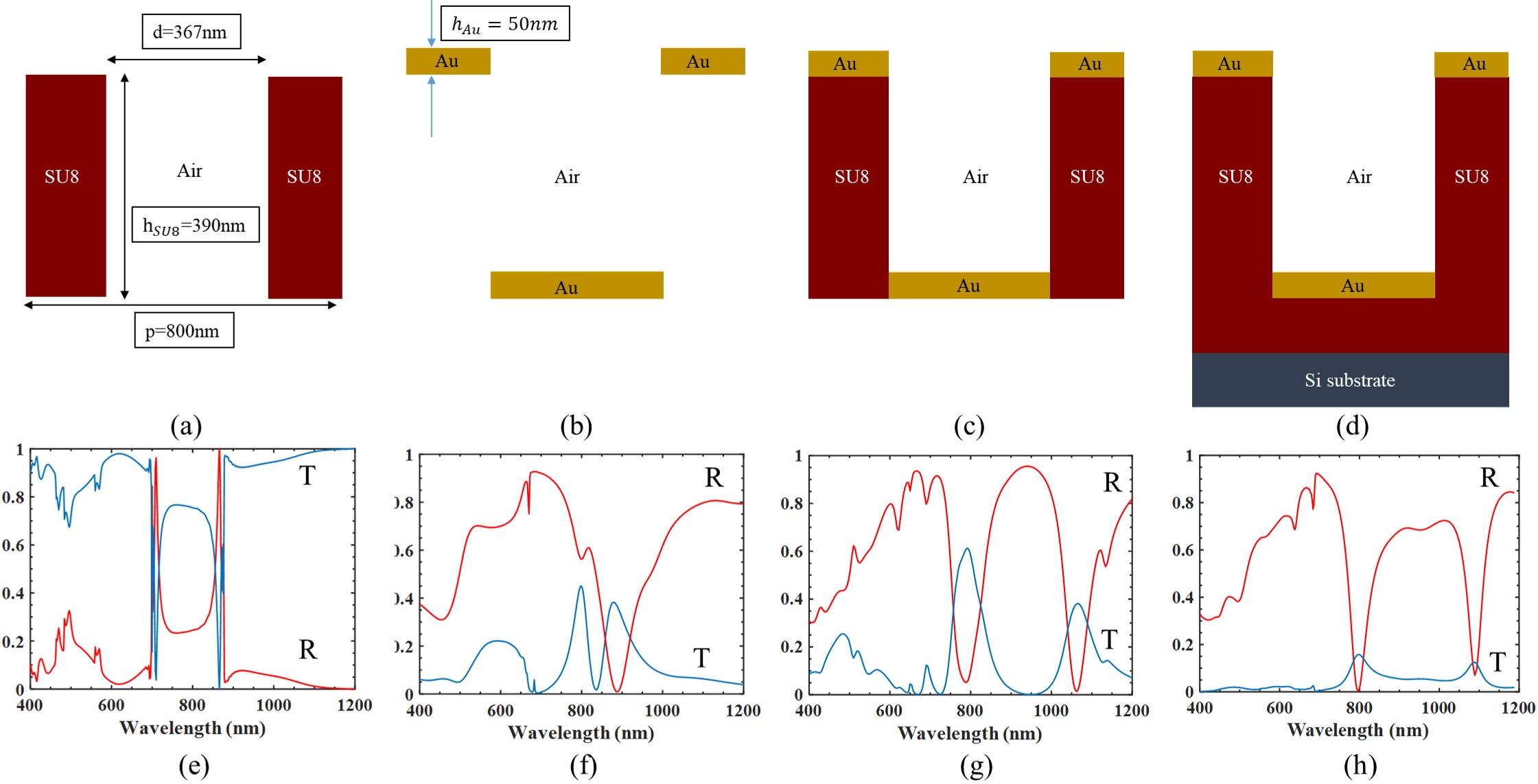}
    \caption{\label{fig2}(a)-(d) Quasi-3D structure and its breakdown into components, as modelled in FDTD simulations (e)-(h) Transmittivity and reflectivity of the corresponding systems}
\end{figure*}

Full 3D finite-difference time-domain (FDTD) simulations employing appropriate periodic boundary conditions \cite{bib24,bib25,bib26} were performed to precisely model the reflection and transmission spectra, as well as the EM field distributions in and around the metal nanostructure (details are provided in the Supporting Information). The simulations were conducted for both TM and TE polarization, however, due to the symmetry of the structure its the resonant properties are independent of the incident light polarization. To investigate the optical response of the complex structure and properly understand the origin of its characteristic resonances, we decompose our structure down into three distinct structures and simulated them individually as follows: a resist-only configuration (Fig.\ref{fig2}(a)), a periodic gold double layer configuration suspended in air (Fig.\ref{fig2}(b)), and the complete hybrid system (Fig.\ref{fig2}(c)). For each case, both the reflectivity and transmittance were modeled, and the corresponding electric field distributions at the critical wavelengths were determined. The PhC is formed by a hexagonal lattice of holes patterned in SU8 polymer, where a single unit cell of the lattice (Fig.\ref{fig1}(c)) is a rectangle of length $p$ and $\sqrt{3}*p$ in the x, and y direction respectively, giving rise to two different resonant wavelengths where the varying lattice parameters along the x and y directions lead to the formation of two distinct resonant wavelengths (as shown in Fig.\ref{fig2}(e)), independent of the incident light polarization. We then simulate a periodic double layer configuration in gold, where the top layer is a 2D array of nano-holes in the gold film, and a 2D array of isolated gold disks in the bottom, fully suspended in air. This second system is characterized by a pure plasmonic resonance (Fig.\ref{fig2}(f)). 

When the two isolated structures are compiled together, we get the response of the characteristic hybrid-plasmonic system, displaying two resonances at 790 nm and 1060 nm (Fig.\ref{fig2}(g)). However, in reality these samples are fabricated on a silicon substrate as well as has an excess layer of SU8 at the bottom of the holes. When they are introduced into the simulation (Fig.\ref{fig2}(d)), the two resonances are still present but the second one is notably attenuated with respect to the previous case (Fig.\ref{fig2}(h)). The resonance at 790 nm is attributed to a surface lattice resonance (SLR), which results from the coupling between the Rayleigh anomaly (RA) and the localized SP resonance (LSPR) and it will be the resonance of interest for our NL studies. SLRs are a pivotal mechanism in optical metasurfaces, yielding spectral responses with ultranarrow linewidths by effectively suppressing radiation losses in metals. This suppression leads to significantly stronger electric field enhancement and localization compared to those of LSPR \cite{bib27,bib28,bib29,bib30,bib31}. The pronounced dips in the transmission spectrum are caused by light confinement within the holes at specific wavelengths, with the resonance positions being highly sensitive to the geometrical parameters of the structure, namely, the period (p), diameter (d), gold layer thickness ($h_{Au}$), and hole depth ($h_{SU8}$) (the tunability of the structure is further explored in the Supporting Information).

\begin{figure*}
    \includegraphics[width=.5\textwidth,center]{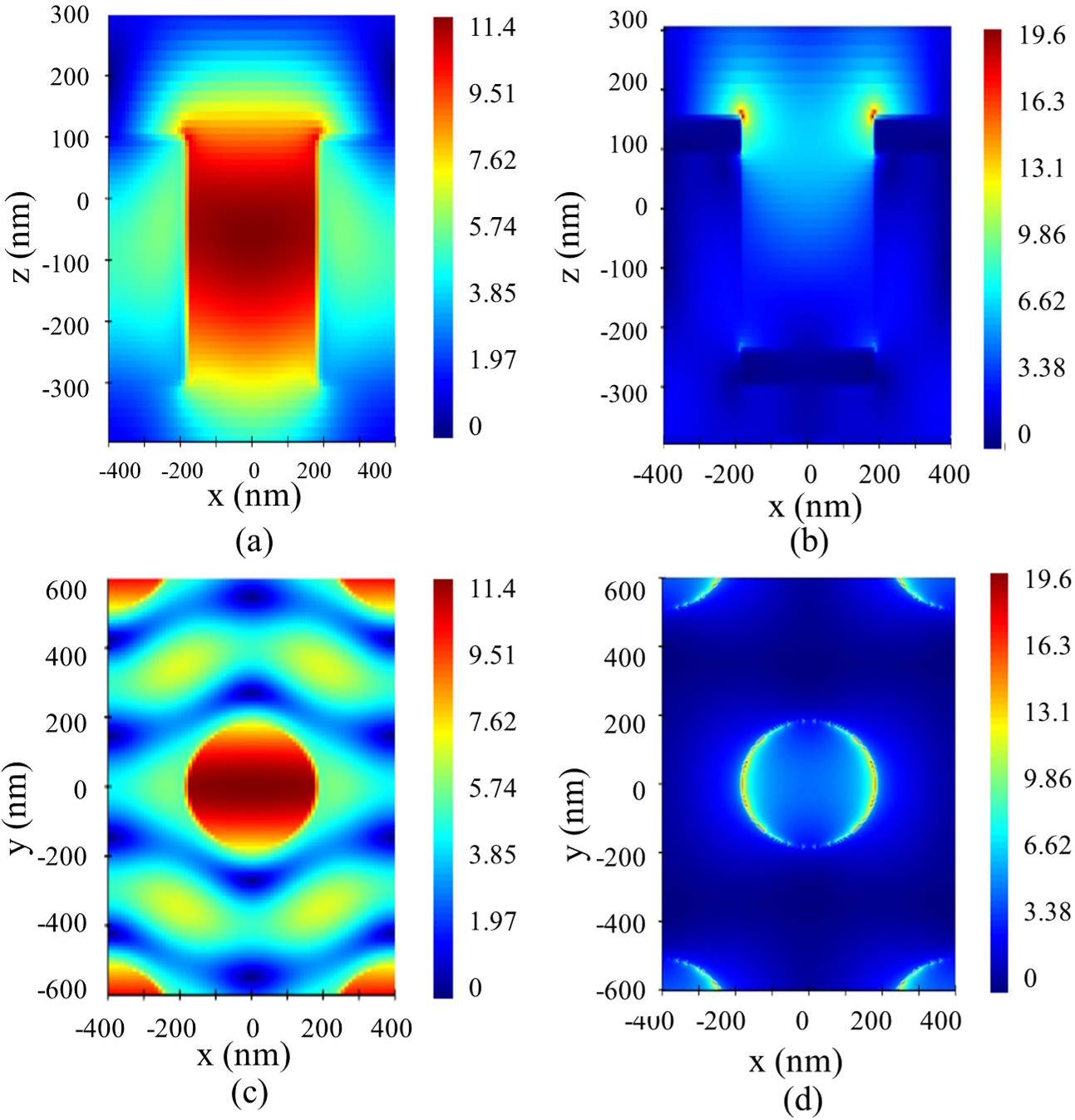}
    \caption{\label{fig3}(a)-(b) Vertical map of E field localization (c)-(d) Horizontal map of E field localization, in the wavelength of resonance. The left column corresponds to the dielectric PhC at $\lambda_{resonance}\sim850 nm$ and The right column corresponds to the Quasi-3D structure at $\lambda_{resonance}\sim790 nm$.}
\end{figure*}

This hybrid metal-dielectric plasmonic resonance not only intensifies the localization of the electric field but also confines it spatially at the metal-air interface. This is evident in the following maps of the E field distribution in the vertical and horizontal cross-sections of the structure, respectively. In case of the resist-only PhC configuration shown in Fig.\ref{fig2}(a), the E field is largely localized in the hole, both in vertical (Fig.\ref{fig3}(a)) and horizontal (Fig.\ref{fig3}(c)) cross-sectional view. On the right side, the quasi-3D structure at its wavelength of resonance $\sim$790 nm, strongly localizes the E field at the edge of the hole, both in vertical (Fig.\ref{fig3}(b)) and horizontal (Fig.\ref{fig3}(d)) cross sectional view. This strong spacial localization at the metal-air interface ultimately promotes stronger light–matter interactions at the metal-air interface, thereby enhancing the NL dynamics at the nanoscale. 

\section{\label{sec:3}Sample fabrication and linear characterization}

\begin{figure*}[h]
    \includegraphics[width=0.9\textwidth,center]{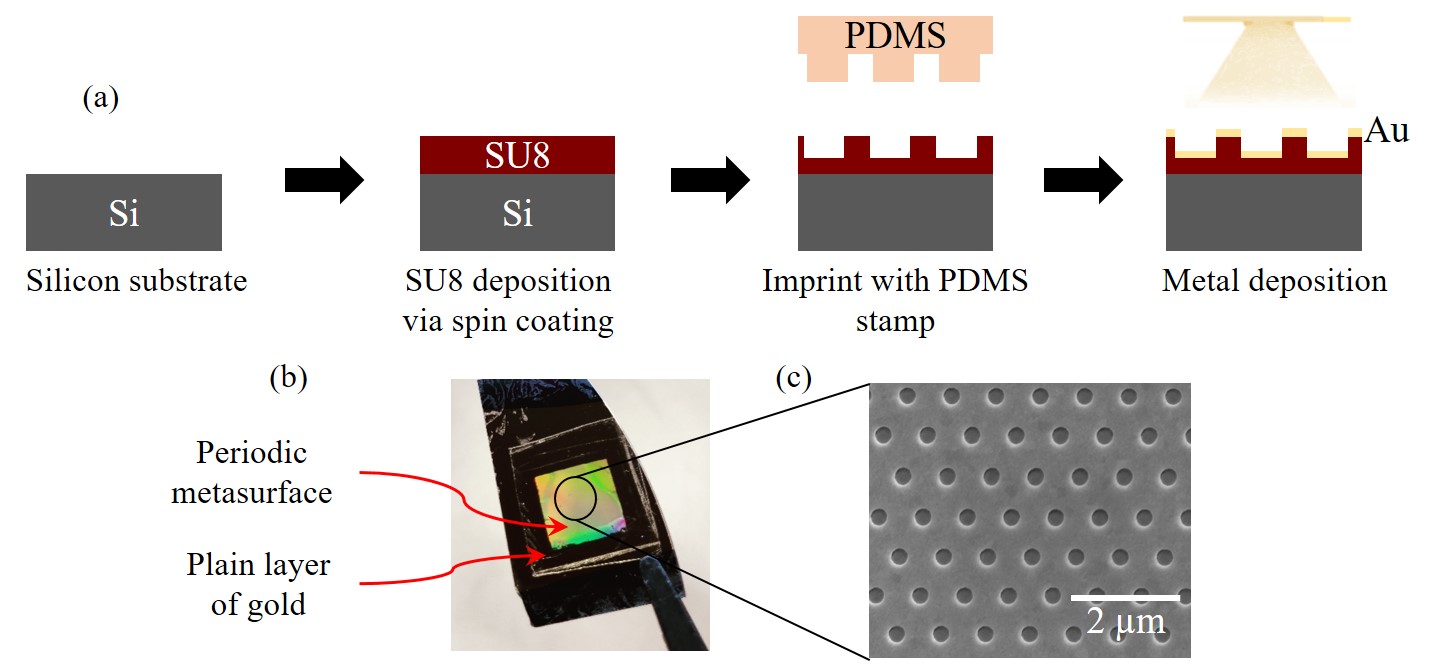}
\caption{\label{fig4} (a) Schematic representation of the fabrication process. (b) Real image of the sample showing the periodically patterned and the unpatterned part (c) Scanning electron microscope (SEM) image of the sample.}
\end{figure*}

We present a simple yet robust fabrication method based on soft-imprinting (see Fig.\ref{fig4})\cite{bib32}. Silicon chips were first cleaned in ultrasonic baths of acetone and isopropyl alcohol for 5 min each. Subsequently, the SU8 photoresist (MicroChem 2000.5) was spin coated at 2000 rpm for 30 s and soft baked at 60$^\circ$C, resulting in a uniform layer approximately 600 nm thick.

The SU8 layer was then patterned using a hybrid soft PDMS–hard PDMS stamp featuring pillars arranged in a hexagonal lattice. Imprinting was performed at 90$^\circ$C under a pressure of 2 bar for 30 s, followed by demolding at 50$^\circ$C. This imprinting process created holes in the resist with a diameter of approximately 367 nm, a depth of around 390 nm, and a center to center distance of 800 nm.

After patterning, the nanostructured resist films were exposed to UV light for 10 min to induce crosslinking, and then post-baked at 150$^\circ$C for 30 min. Finally, a 50 nm gold layer was deposited using a thermal evaporator (MBraun). This directional deposition coated both the top surface and the recessed regions with gold, while leaving the nanohole sidewalls uncoated, thereby completing the fabrication of the plasmonic nanostructures.

\section{\label{sec:4}Experimental setup for the measurement of THG}

\begin{figure*}
\includegraphics[width=.9\textwidth,center]{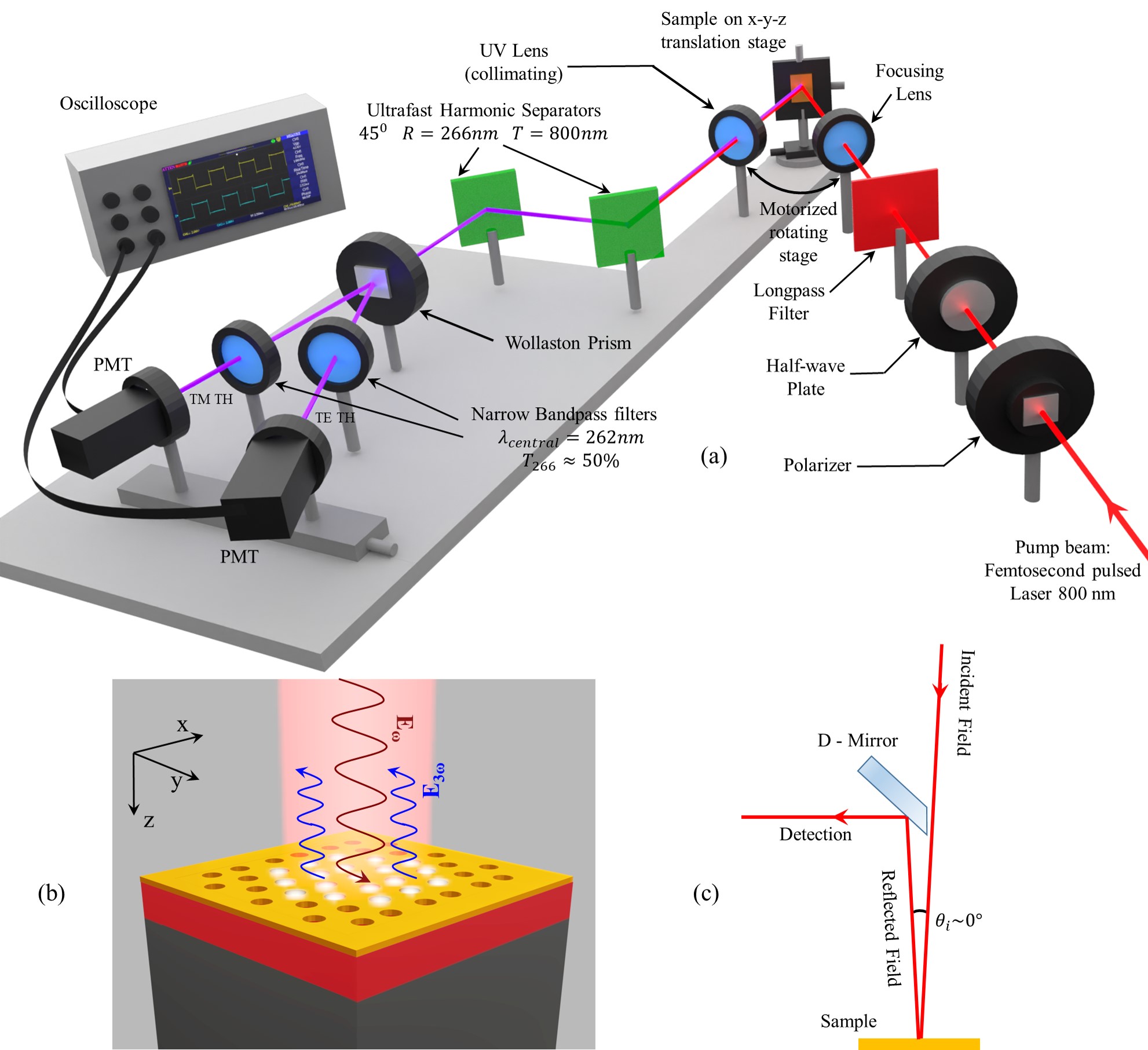}
\caption{\label{fig5} (a) Schematics of the experimental set-up used to measure reflected third harmonic signals generated by the gold grating as a function of angle of incidence, polarization, and incident wavelength. (b) Schematic representation of to E field localization and THG at normal incidence (c)setup modification to detect the THG at normal or near normal incidence.}
\end{figure*}

To efficiently detect the THG from the nanostructures, we constructed the setup shown in Fig.\ref{fig5}(a). The setup allows us to measure the efficiency of power conversion from the incident fundamental (FF) to the generated TH frequency, defined as $\eta_{TH}=\frac{P_{TH}}{P_{FF}}$, ranging from $10^{-8}$ down to $10^{-13}$. The light source is a tunable Ti:Sapphire femtosecond laser operating around 800 nm with a pulse duration of approximately 170 fs (at full-width half maxima), a repetition rate of 76 MHz, and a continuous-wave (CW) average power of 1 W. A half-wave plate at the laser output controls the polarization (TE or TM) of the FF beam, while a long-pass filter eliminates any spurious NL signals generated by other optical components along the setup. The pump beam is focused on the sample using a plano-convex lens with a focal length of 10 cm to achieve intensities between 1 and 4 GW/cm$^2$. The gold samples are mounted on an x–y–z translation stage, which is fixed to a motorized rotary support that adjusts the angle of incidence with a precision of $0.2^\circ$.

Fig.\ref{fig5}(a) also illustrates the TH detection arm, which consists of an anti-reflection coated (UV-enhanced) collimating lens, two mirrors acting as “ultrafast harmonic separators” (reflecting the THG wavelength while allowing the pump wavelength to pass), and a narrow bandpass filter that transmits primarily around the TH frequency while nearly completely attenuating the pump radiation. A Wollaston polarizer then separates the TM and TE polarization components of the TH signal for simultaneous detection, and a photomultiplier tube (PMT) is used to detect the harmonic signals. Additionally, a chopper placed immediately after the input laser modulates the beam with a square-wave profile to facilitate signal detection.

A comprehensive calibration procedure accounting for the transmission characteristics of each optical component as well as the wavelength-dependent responsivity of the PMT allows us to accurately estimate the conversion efficiencies as the ratio of the TH energy to the energy of the incident FF beam. The entire detection system, which is the most critical part of the setup, is mounted on a rotary platform that enables a precise setting of the detection angle for reflection measurements. A detailed alignment and calibration process was also carried out for this detection setup. To calculate the conversion efficiencies, the transmittance of the filters, lenses, and mirrors, as well as the responsivity of the detectors, placed after the sample, were carefully accounted for.

The gold nanohole structure exhibits its sharpest resonance at normal incidence (Fig.\ref{fig5}(b)), while increasing the angle broadens the resonance and reduces its depth (Supplementary Fig. S5). In our previous study on SHG and THG from 20 nm and 70 nm gold nanolayers, we observed that the emission from the bare metal displayed an angular dependence-favoring maximum THG at normal incidence \cite{bib3}. Consequently, our current work focuses on the experimental detection of the THG emitted from the periodic quasi-3D gold structure, induced by the plasmonic resonant field localization, and compares it to that of a 50 nm thick gold layer, to precisely quantify the THG enhancement.

To capture the TH signal at normal or near-normal incidence while retaining the core components of the setup, we modified our detection system by incorporating a D-shaped UV-enhanced mirror. As depicted in Fig.\ref{fig5}(c), this mirror is designed to operate at a very small angle, ensuring that the path of incident light remains undisturbed. The D mirror performs best at 45$^\circ$, where it maximizes the reflection of the UV signal and minimizes the reflection of the 800 nm pump. During the calibration process and the precise calculation of TH conversion efficiencies, the reflection characteristics of the D mirror were carefully accounted for.

\section{\label{sec:5}Results and Discussion}

\begin{figure}
\includegraphics[width=.5\textwidth,center]{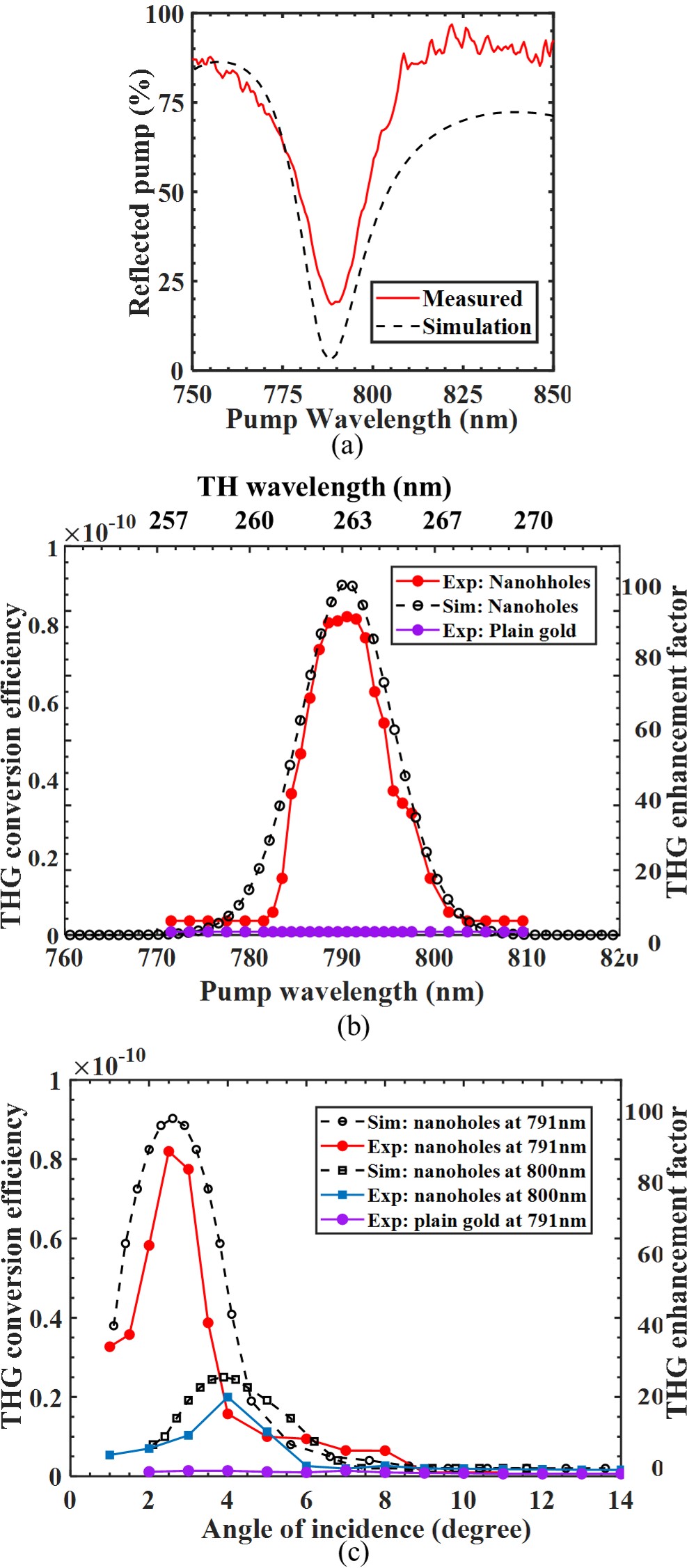}
\caption{\label{fig6}(a) Plasmonic resonance centered around 790 nm at normal incidence for the fundamental beam; (b) Simulated and experimentally detected spectral profile of the resonant THG enhancement (left y-axis) and THG conversion efficiency (right y-axis) around the wavelength of resonance (c) Simulated and experimentally detected angular profile of the resonant THG enhancement (left y-axis) and THG conversion efficiency (right y-axis) at several near zero angles of incidence. }
\end{figure}

Fig.\ref{fig6}(a) displays a narrow plasmonic resonance centered at 790 nm. The theoretical spectral response in reflection was computed using full 3D FDTD simulations \cite{bib24,bib25,bib26}, and the experimental spectrum-recorded at near-normal incidence with TM-polarized light and shown as the red continuous curve-is in excellent agreement with the theoretical prediction (black dashed curve).

The gold coating is thick enough that the transmission through the sample is nearly zero, as confirmed by both simulation and experiment. The depth and position of the resonance depend critically on several geometrical parameters, including the period (p), diameter (d), and the depth of the holes ($h_{SU8}$), as well as the thickness of the deposited gold layer ($h_{Au}$). Notably, the experimental resonance linewidth is broader than the simulated one, likely because the excitation source deviates slightly from an ideal plane wave as it propagates through free space before reaching the sample as well as the imperfections in the sample post fabrication.

The theoretical prediction, also calculated via FDTD \cite{bib33,bib34,bib35} and represented by the black dotted curve, shows a maximum enhancement factor of 108 (Fig.\ref{fig6}(b)), with respect to the THG from a plain layer of gold of $h_{Au} \sim 50 nm$. The conversion efficiency is highly sensitive to the peak power, pulse width, and shape of the pump pulse. In our NL simulations, the source amplitude was defined to match the average pump laser intensity recorded experimentally: Amplitude =$\sqrt{\frac{I_{peak}}{\epsilon_0 c}}\sim 10^8  V/m$ with the source pulse duration set to $\sim$170 fs. The overall conversion efficiency $\eta_{(SH/TH)}$  depends on the average excitation intensity and the third-order NL susceptibility $\chi^{(3)}$ of gold. To account for this, a wavelength-dependent $\chi^{(3)}$  database for bulk gold was incorporated into the FDTD database. Fig.\ref{fig6}(b) also illustrates the spectral response of the resonant THG emission, which is corroborated by our experimental measurements taken at near-normal incidence by tuning the Ti:Sapphire laser pump around the resonance wavelength. The experimentally retrieved resonant THG from the quasi-3D structure, represented by red dots, indicates a maximum enhancement factor of 98, with respect to the experimentally detected THG from the unpatterned layer of gold (Fig.\ref{fig6}(b) purple curve) under the same laboratory conditions.

Additional measurements were conducted to assess the resonant behavior of the THG as the angle of incidence approaches normal incidence. Fig.\ref{fig6}(c) presents three sets of experimental data: THG efficiency from a plain 50 nm gold layer (violet curve), THG efficiency from the quasi-3D structure when illuminated by the pump at 800 nm (blue curve), and THG efficiency from the quasi-3D structure when illuminated by the pump at 791 nm (red curve). As observed, we tend to approach the resonant THG peak as we approach normal incidence. Experimentally we are able to precisely retrieve the resonant THG peak at 2$^\circ$ angle of incidence, which shows a maxima at 791 nm, slightly red shifted from the 790 nm resonance at strictly normal incidence. The simulated curve of the resonant THG peak (black dotted curve with round markers) is slightly wider and shows a smoother transition into and out of the resonance, while the red curve shows a narrower, more abrupt peak. This is due to the diffraction and losses in the real sample, and as soon as we move out of the resonance, we abruptly lose the quality of the resonance. Experimentally we also retrieve the resonant THG curve for a further red-shifted pump wavelength of 800 nm, which shows a maxima at 4$^\circ$ angle of incidence (blue curve) and its corresponding simulated curve represented by black dotted curve with square markers. In all four curves, as the angle of incidence increases and we move out of the resonance, the THG efficiency decreases significantly, while the THG from plain layer of gold shows no such significant transition.

The exceptional sensitivity of the nanohole structure, which also makes it an excellent sensing platform \cite{bib20,bib21,bib22,bib23}, seems to limit its THG efficiency, compared to the simple full metal structure (example a 2D gold grating that we studied \cite{bib8}). The reason, as we discuss now, contributes to our understanding of these complex hybrid structures under resonant illumination conditions, as well as opens up the possibility to further improve the sample configuration for much larger NL generation. In configurations designed to harness the third-order nonlinearities of gold, a gold grating typically offers robust electric field confinement at the trench corners \cite{bib8}, substantially enhancing light–matter interactions. This is primarily because THG relies on the field penetration depth to exploit the bulk third-order nonlinearity ($\chi^{(3)}$) of the metal. Although plasmonic resonances intensify the electric field at the surfaces and corners of the holes or disks, most of the field still extends into the interior of the holes(Fig.\ref{fig3}(d)). Consequently, the maximum field intensity does not effectively overlap with the gold, resulting in significantly lower THG signals.

Nevertheless, varying the geometric parameters-such as using square or round holes, adopting square or hexagonal lattice periodicities, and modifying the hole depth and diameter-can lead to different outcomes. This sensitivity and tunability of the plasmonic resonance open up broad possibilities for the design, characterization, and fabrication of these structures, especially in sensing applications (for example, detecting a biomolecule in a solvent).

\begin{figure*}[h]
\includegraphics[width=.9\textwidth,center]{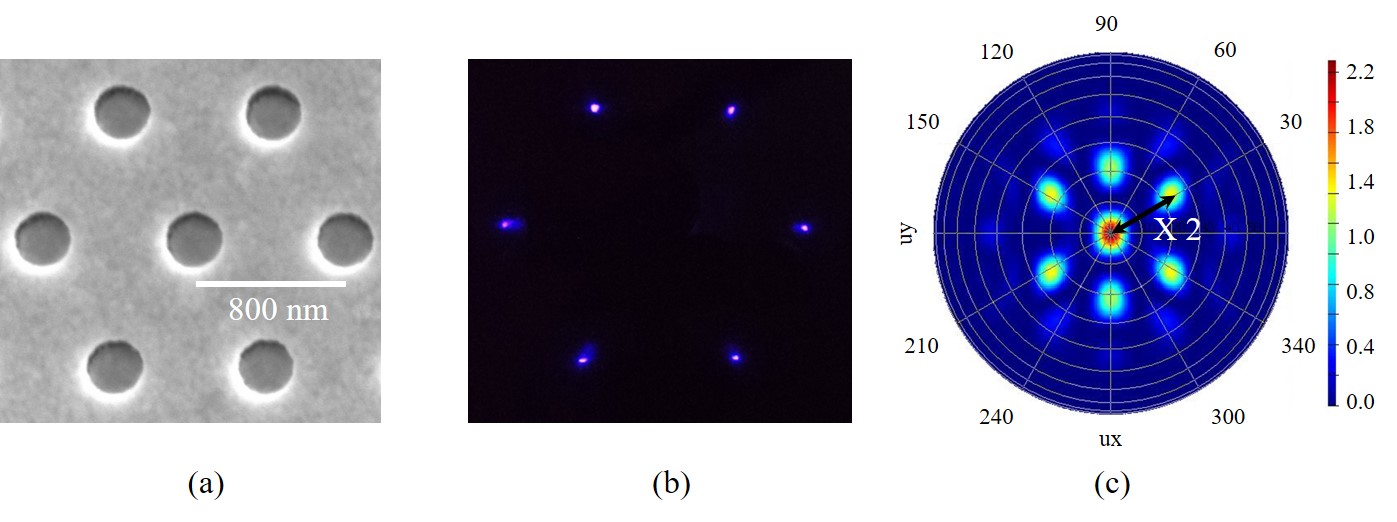}
\caption{\label{fig7} (a) Top-view SEM image of the hexagonal lattice sample (b) Experimental verification of the diffraction pattern of the generated higher harmonic (c) Simulated diffraction pattern of the generated higher harmonic. The amplitude of the central spot is two times the amplitude of the first order diffracted spots.}
\end{figure*}

Fig.\ref{fig6} shows the measured enhancement of the THG signal emitted at the angle of reflection. However, the periodicity of the structures can lead to constructive interference of the generated TH in specific directions, producing diffraction orders and making it challenging to capture the total emitted signal. In our structure, the pump wavelength is effectively absorbed, but the generated third harmonic at 263 nm, which is smaller than the periodicity of the structure, is consequently diffracted. Fig.\ref{fig7}(a) presents a SEM image of the sample and Fig.\ref{fig7}(b) presents the corresponding hexagonal diffraction pattern, obtained when the structure is artificially illuminated with the detection wavelength. In previous work with a diffraction gold grating, we confirmed that these diffracted harmonics were emitted at the angles predicted by linear diffraction of the generated harmonics \cite{bib8}. Taking this into account, the FDTD calculations indicate that normal-incidence 263 nm light from our quasi-3D structure is diffracted at an angle of 56$^\circ$. These calculations also provide the intensity ratio of the central spot to the first diffracted spot, which, for our specific hexagonal pattern, is a factor of 2 as shown in Fig.\ref{fig7}(c). When the total THG is considered - combining the zero-order specular emission with all diffracted orders - the overall enhancement factor is approximately 400 times greater than that of a bare gold layer.

The successful detection of extremely low-intensity harmonic generation deep in the UV range indicates that further structural optimization could significantly boost process efficiencies. Since the third-order nonlinearity $\chi^{(3)}$ of gold is an intrinsic bulk property, EM field penetration into the metal is essential to enhance light–matter interactions. In our current structure, however, the electric field is predominantly confined to the metal-air interface, with minimal penetration into the metal. This can be improved by strategically positioning NL material nanoparticles in the nano-gap hotspots of dimers \cite{bib36,bib37,bib38} or by embedding HHG materials within gold shells or rings as demonstrated in \cite{bib39}. These configurations achieve efficient THG by ensuring an optimal spatial overlap between the plasmonic enhancement and the NL material. Additionally, such advanced experimental systems enable a detailed analysis of the individual contributions from the polymer and the metal, thereby shedding light on the origins of the enhanced nonlinear response of the hybrid system.

\begin{figure*}
\includegraphics[width=.9\textwidth,center]{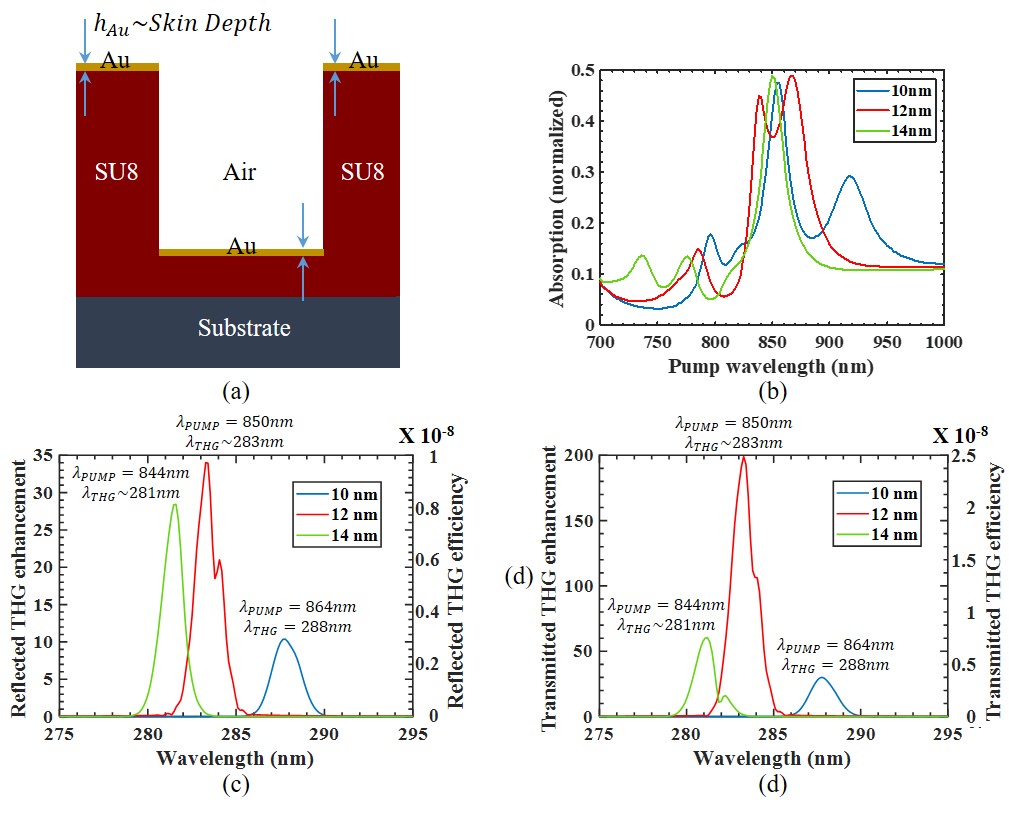}
\caption{\label{fig8} (a) Quasi-3D structure with ultra-thin gold, as modelled in FDTD simulations (side view) (b) spectral resonances corresponding to the absorption maxima (c) Transmitted and (d) Reflected THG efficiency from the ultra-thin Quasi-3D gold structures. The figures represent the resonant THG spectrum and the corresponding enhancement factors with respect to equivalent thickness of unpatterned gold, as simulated from the structures with $h_{Au}$ of 10, 12 and 14 nm.}
\end{figure*}

Alternatively, one could explore the “ultrathin metal” regime, where the metal thickness is of the order of its skin depth - ensuring that the spectral position of the resonantly excited SP waves aligns with the absorption maximum \cite{bib40, bib41,bib42,bib43}. This distinction between thick and ultrathin metal regimes is pivotal, as it directly affects the structure’s ability to absorb and localize light. In the following section, the Quasi-3D structure is remodeled with $h_{Au}$ comparable to or less than the skin depth of the metal (Fig.\ref{fig8}(a)), in stark contrast to the widely studied configurations of thicker layers. Theoretical predictions for light propagation in a quasi-3D structure with an ultrathin gold layer reveal an extraordinary absorptivity of up to $50\%$. While previous studies on grooved structures with thick or semi-infinite metal layers infer SP excitation by monitoring spectral resonances in reflection or transmission, ultrathin structures exhibit resonantly excited SP waves that coincide with the absorption maximum. This resonant absorption greatly enhances the EM field concentration and light penetration inside and throughout the bulk of the metal layer, and takes advantage of the bulk third-order nonlinearity ($\chi^{(3)}$) of the metal, thus proving to be a very promising source of THG.

In structures with ultra-thin layers of gold, NL generation is enhanced through two primary mechanisms. First, a gold nanolayer with a thickness of 10–15 nm already yields an order of magnitude larger TH signal than a 50 nm layer, because a 50 nm layer essentially acts as a mirror, reflecting much of the incident light and confining THG to reflection only. In contrast, an ultra-thin layer absorbs light more efficiently, harnessing the enhanced field concentration and penetration throughout its entire bulk to produce THG in both transmission and reflection. Secondly, periodic patterning of these ultra-thin metals shows their corresponding strong plasmonic resonance, which then furthermore boosts the EM field localization in the wavelength of resonant absorption.

The transition from the thick metal to the ultra-thin metal regime is gradual and can be monitored by varying $h_{Au}$, while keeping the other geometrical parameters of the structure fixed. A sweep of $h_{Au}$ from 5 to 20 nm shows that the plasmonic resonance corresponding to the resonant absorption in the 800 nm range reaches its maximum for $h_{Au}$ of 10, 12 and 14 nm (Fig.\ref{fig8}(b)). Next we simulate THG from each structure, considering that the structures are illuminated by a pump laser tunable around the structure's own wavelength of resonance, and the other characteristics of the pump laser fixed as mentioned in section\ref{sec:4}. Fig.\ref{fig8}(c) and Fig.\ref{fig8}(d) represent our theoretical predictions of the spectral profile of the resonant THG from the structures in transmission and in reflection respectively, corresponding to the quasi-3D structure with $h_{Au}$ as of 10nm (green curve), 12nm (red curve) and 14nm (blue curve), considering the other geometrical parameters of the structure fixed and considering in this case a transparent substrate. In each graph, the left y-axis respresents the resonant THG enhancement factor with respect to that of a plain gold layer of the corresponding thickness, and the right y-axis represents the THG conversion efficiencies, around the wavelength of resonance.

\section{\label{sec:6}Conclusion}

In this paper, we present and study resonant THG from a quasi-3D gold hybrid plasmonic metasurface. We performed a set of
simulations to understand and tune the linear and nonlinear resonant properties of the structure at or near zero angle of incidence,
as well as a series of experiments to corroborate those results regarding the shape, spectral width, and peak amplitude of the
spectral response. Our sophisticated detection system and calibration procedures enable accurate experimental verification of
the THG efficiencies emitted from the resonant quasi-3D gold structure, as well as its enhancement with respect to simple plain
layer of gold. When the total THG efficiency - defined as the sum of the zeroth and first diffraction orders - is scaled to the THG
reference signal from an unpatterned region of the same gold sample (measured under identical conditions), an enhancement
factor of approximately 400 is obtained. Numerical modeling quantitatively captures the observed optical properties of these
architectures and reveals the complex electromagnetic EM field distributions around their multilevel nanostructured
features, which in turn drive the enhanced NL interactions at the metal surface. It is important to note that unlike
convensional observations, we are measuring the real efficiencies of conversion from the pump to the third harmonic,
directly in terms of incident and generated power. Secondly, our hybrid quasi-3D metasurface already proves to be more
efficient in light conversion than more complicated structures, such as \cite{bib10} and \cite{bib13}, each of which is technologically challenging to
fabricate, however show only one order of magnitude enhancement.

Substituting our 50 nm thick Au layer by ultrathin gold layers would open the door to exploiting THG emission in
transmission and its further enhancement. To achieve this, the hybrid structure must be fabricated on a substrate that is
transparent to both the FF wavelength and the THG wavelength. However, finding a suitable substrate is challenging because
most optical components/materials are highly absorptive in the deep UV range and is part of future work.

This study challenges the notion that a more complex metasurface inherently produces higher nonlinearity. Actually In
reality, the successful generation and detection of higher-order harmonics depend on a detailed understanding of the plasmonic structure, its resonant properties, and the NL electron dynamics at the nanoscale. Such insights are fundamental for developing coherent and efficient nanoscale UV/DUV/EUV light sources that facilitate miniaturization and integration into
portable devices. Moreover, many photonic devices require operation at small angles of incidence. The precise engineering
of metasurface geometry, combined with modern fabrication methods, promises highly tailored plasmonic structures capable
of producing compact high-frequency light sources. These structures can be embedded with NL components to achieve
beam deflection, metalensing, meta-holography, and even the generation of orbital-angular-momentum beams.

Our experimental findings, corroborated by rigorous electromagnetic simulations, establish this volumetric metasurface
as a robust, real-world platform for deep-UV light sources, nonlinear sensors, and emerging quantum photonic applications.
This genuinely three-dimensional architecture paves the way for reconfigurable, multi-polarization nanophotonic devices
that fully leverage the spatial richness of light–matter interactions at the nanoscale. Ultimately, multifunctional plasmonic
materials that enable dynamic manipulation of light properties-such as amplitude and polarization-at the nanoscale are critical
for the advancement of future all-optical nanotechnologies based on active photonics.

\bibliography{sample}

@PREAMBLE{
 "\providecommand{\noopsort}[1]{}" 
 # "\providecommand{\singleletter}[1]{#1}%" 
}

@BOOK{bib1,
   author       = {R. W. Boyd},
   year         = 2003,
   title        = {Nonlinear Optics (3rd ed.)},
   publisher    = {Academic Press}
}

@BOOK{bib2,
   author       = {Y. R. Shen},
   year         = 2003,
   title        = {Principles of Nonlinear Optics},
   publisher    = {Wiley-Interscience}
}

@ARTICLE{bib3,
   author       = " L. Rodríguez-Suné and J. Trull and C. Cojocaru and N. Akozbek and D. De Ceglia and M. A. Vincenti and M. Scalora",
    title       = "Harmonic generation from gold nanolayers: bound and hot electron contributions to nonlinear dispersion",
   year         = "2021",
   journal      = "Optics Express",
   volume       = "29",
   pages        = "8581",
}

@ARTICLE{bib4,
   author       = "M. Scalora and K. Hallman and S. Mukhopadhyay and S. Pruett and D. Zappa and E. Comini and D. de Ceglia and M. A. Vincenti and N. Akozbek and J. Trull and C. Cojocaru", 
    title       ="Below the surface: Unraveling the intricacies of the nonlinear optical properties of aluminum through bound electrons",
   year         = "2024", 
   journal      = "APL Photonics", 
   volume       = "9", 
   pages        = "086108",
}

@BOOK{bib5,
   author       = "Stefan A. Maier", 
    title       ="Plasmonics: Fundamentals and Applications",
   year         = "2007",
   publisher    = {Springer}
}

@ARTICLE{bib6,
   author       = "M. Kauranen and A. Zayats", 
    title       ="Nonlinear plasmonics ",
   year         = "2012", 
   journal      = "Nature Photonics ", 
   volume       = "6", 
   pages        = "737–748",
}

@ARTICLE{bib7,
   author       = "S. Kim and T. In Jeong and J. Park and M. F. Ciappina and S. Kim ", 
    title       ="Recent advances in ultrafast plasmonics: from strong field physics to ultraprecision spectroscopy",
   year         = "2022", 
   journal      = "Nanophotonics", 
   volume       = "11", 
   pages        = "2393-2431",
}

@ARTICLE{bib8,
   author       = "S. Mukhopadhyay and L. Rodriguez-Suné and C. Cojocaru and M. A. Vincenti and K. Hallman and G. Leo and M. Belchovski and D. de Ceglia and M. Scalora and J. Trull ", 
    title       ="Three orders of magnitude enhancement of second and third harmonic generation in the visible and ultraviolet ranges from plasmonic gold nanogratings",
   year         = "2023", 
   journal      = "APL Photonics", 
   volume       = "8", 
   pages        = "046108",
}

@ARTICLE{bib9,
   author       = "A. Falamas and V. Tosa and C. Farcau", 
    title       ="Hybrid architectures made of nonlinear-active and metal nanostructures for plasmon-enhanced harmonic generation",
   year         = "2019", 
   journal      = "Optical Materials", 
   volume       = "88", 
   pages        = "653-666",
}

@ARTICLE{bib10,
   author       = "S. Chen and W. Wong and Y. Pun and  K. Cheah and G. Li", 
    title       = "Surface plasmon-enhanced third harmonic generation from gold–polymer hybrid plasmonic crystal",
   year         = "2013", 
   journal     =  "Advanced Optical Materials", 
   volume       = "1", 
   pages        = "522-526",
}

@ARTICLE{bib11,
   author       = "F. Ren and X. Wang and Z. Li and J. Luo and S. H. Jang and A. K. Jen and A. X. Wang", 
    title       ="Enhanced third harmonic generation by organic materials on high-Q plasmonic photonic crystals",
   year         = "2014", 
   journal      = "Optics Express", 
   volume       = "22", 
   pages        = "20292–20297",
}

@ARTICLE{bib12,
   author       = "G. Albrecht and M. Hentschel and S. Kaiser and H. Giessen", 
    title       ="Hybrid organic-plasmonic nanoantennas with enhanced third-harmonic generation",
   year         = "2017", 
   journal      = "ACS Omega", 
   volume       = "2", 
   pages        = "2577–2582",
}

@ARTICLE{bib13,
   author       = "O. Bar-Elli and E. Grinvald and N. Meir and L. Neeman and D. Oron", 
    title       ="Enhanced third-harmonic generation from a metal/semiconductor core/shell hybrid nanostructure",
   year         = "2015", 
   journal      = "ACS Nano", 
   volume       = "9", 
   pages        = "8064–8069",
}

@ARTICLE{bib14,
   author       = "S. Kim and J. Jin and Y. J. Kim and I. Y. Park and Y. Kim and S. W. Kim", 
    title       ="High-harmonic generation by resonant plasmon field enhancement",
   year         = "2008", 
   journal      = "Nature", 
   volume       = "453", 
   pages        = " 757–760 ",
}

@ARTICLE{bib15,
   author       = "M. Celebrano and X. Wu and M. Baselli and S. Großmann and P. Biagioni and A. Locatelli and C. De Angelis and G. Cerullo and R. Osellame and B. Hecht and L. Duò and F. Ciccacci and M. Finazzi ", 
    title       ="Mode matching in multi resonant plasmonic nanoantennas for enhanced second harmonic generation",
   year         = "2015", 
   journal      = "Nature Nanotechnology ", 
   volume       = "10", 
   pages        = "412–417",
}

@ARTICLE{bib16,
   author       = "M. Baselli and A. L. Baudrion and L. Ghirardini and G. Pellegrini and E. Sakat and L. Carletti and A. Locatelli and C. De Angelis and P. Biagioni and L. Duò and M. Finazzi and P. M. Adam and M. Celebrano", 
    title       ="Plasmon-Enhanced Second Harmonic Generation: from Individual Antennas to Extended Arrays",
   year         = "2016", 
   journal      = "Plasmonics ", 
   volume       = "12", 
   pages        = "1595–1600",
}

@ARTICLE{bib17,
   author       = "J. Alberti and H. Linnenbank and S. Linden and  Y. Grynko and J. Förstner ", 
    title       ="The role of electromagnetic interactions in second harmonic generation from plasmonic metamaterials",
   year         = "2016", 
   journal      = " Applied Physics B ", 
   volume       = "122", 
   pages        = "45",
}

@ARTICLE{bib18,
   author       = "Y. Wang and D. Wei and Y. Zhu and X. Huang and X. Fang and W. Zhong and Q. Wang and Y. Zhang and Min Xiao", 
    title       ="Conversion of the optical orbital angular momentum in a plasmon-assisted second-harmonic generation",
   year         = "2016", 
   journal      = "Applied Physics Letters", 
   volume       = "103", 
   pages        = "081105",
}

@ARTICLE{bib19,
   author       = "R. Kolkowski and J. Szeszko and B. Dwir and E. Kapon and J. Zyss", 
    title       ="Non-centrosymmetric plasmonic crystals for second-harmonic generation with controlled anisotropy and enhancement",
   year         = "2016", 
   journal      = "Laser I\& Photonics Reviews", 
   volume       = "10", 
   pages        = "287-298",
}

@ARTICLE{bib20,
   author       = "D. Chanda and K. Shigeta and T. Truong and E. Lui and A. Mihi and M. Schulmerich and P. V. Braun and R. Bhargava and J. A. Rogers", 
    title       ="Coupling of plasmonic and optical cavity modes in quasi-three-dimensional plasmonic crystals",
   year         = "2011", 
   journal      = "Nature Communications", 
   volume       = "2", 
   pages        = "479",
}

@ARTICLE{bib21,
   author       = "M. E. Stewart and N. H. Mack and V. Malyarchuk and J. A. N. T. Soares and T. Lee and S. K. Gray and R. G. Nuzzo and J. A. Rogers", 
    title       ="Quantitative multispectral biosensing and 1D imaging using quasi-3D plasmonic crystals",
   year         = "2006", 
   journal      = "Proc. Natl. Acad. Sci. U.S.A.", 
   volume       = "103", 
   pages        = "17143-17148",
}

@ARTICLE{bib22,
   author       = "S. K. Verma and S. K. Srivastava", 
    title       ="Giant Extra-Ordinary Near Infrared Transmission from Seemingly Opaque Plasmonic Metasurface: Sensing Applications",
   year         = "2022", 
   journal      = "Plasmonics ", 
   volume       = "17", 
   pages        = "653–663",
}

@ARTICLE{bib23,
   author       = "M. Zhang and M. Lu and C. Ge and B. T. Cunningham", 
    title       ="Plasmonic external cavity laser refractometric sensor",
   year         = "2014", 
   journal      = "Opt. Express", 
   volume       = "22", 
   pages        = "20347-20357",
}

@ARTICLE{bib24,
   author       = "N. B. Khairulazdan and P. S. Menon and A. R. Md. Zain and D. D. Berhanuddin", 
    title       ="Optimization of photonic crystal structure by FDTD method to improve the light extraction efficiency in silicon",
   year         = "2022", 
   journal      = "Chalcogenide Letters        ", 
   volume       = "19", 
   pages        = "493 - 501",
}

@Inbook{bib25,
author="Fomberg, Bengt",
editor="Ainsworth, Mark
and Davies, Penny
and Duncan, Dugald
and Rynne, Bryan
and Martin, Paul",
title="Some Numerical Techniques for Maxwell's Equations in Different Types of Geometries",
bookTitle="Topics in Computational Wave Propagation: Direct and Inverse Problems",
year="2003",
publisher="Springer Berlin Heidelberg",
address="Berlin, Heidelberg",
pages="265--299",
}

@article{bib26,
author = {Jun-Whee Kim and Ji-Hyang Jang and Min-Cheol Oh and Jin-Wook Shin and Doo-Hee Cho and Jae-Hyun Moon and Jeong-Ik Lee},
journal = {Opt. Express},
number = {1},
pages = {498--507},
publisher = {Optica Publishing Group},
title = {FDTD analysis of the light extraction efficiency of OLEDs with a random scattering layer},
volume = {22},
month = {Jan},
year = {2014},
}

@ARTICLE{bib27,
   author       = "M. Kataja and T. K. Hakala and A. Julku and M. J. Huttunen and S. van Dijken and P. Törmä", 
    title       ="Surface lattice resonances and magneto-optical response in magnetic nanoparticle arrays",
   year         = "2015", 
   journal      = "Nature Communications", 
   volume       = "6", 
   pages        = "7072",
}

@ARTICLE{bib28,
   author       = "S. Hamdad and A. T. Diallo and M. Chakaroun and A. Boudrioua ", 
    title       ="The role of Rayleigh anomalies in the coupling process of plasmonic gratings and the control of the emission properties of organic molecules",
   year         = "2022", 
   journal      = "Scientific Reports", 
   volume       = "12", 
   pages        = "3218",
}

@ARTICLE{bib29,
author = {Kravets, V. G. and Kabashin, A. V. and Barnes, W. L. and Grigorenko, A. N.},
title = {Plasmonic Surface Lattice Resonances: A Review of Properties and Applications},
journal = {Chemical Reviews},
volume = {118},
number = {12},
pages = {5912-5951},
year = {2018},
}

@ARTICLE{bib30,
   author = {Quoc Trung Trinh and Sy Khiem Nguyen and Dinh Hai Nguyen and Gia Khanh Tran and Viet Hoang Le and Hai-Son Nguyen and Quynh Le-Van},
journal = {Opt. Lett.},
number = {6},
pages = {1510--1513},
publisher = {Optica Publishing Group},
title = {Coexistence of surface lattice resonances and bound states in the continuum in a plasmonic lattice},
volume = {47},
month = {Mar},
year = {2022},
}

@ARTICLE{bib31,
  author = {Zhou, Wei and Hua, Yi and Huntington, Mark D. and Odom, Teri W.},
title = {Delocalized Lattice Plasmon Resonances Show Dispersive Quality Factors},
journal = {The Journal of Physical Chemistry Letters},
volume = {3},
number = {10},
pages = {1381-1385},
year = {2012},
}

@ARTICLE{bib32,
   author = "C. Matricardi and J. L. Garcia-Pomar and P. Molet and L. A. Pérez and M. I. Alonso and M. Campoy-Quiles and A. Mihi",
title = "High-Throughput Nanofabrication of Metasurfaces with Polarization-Dependent Response",
journal = "Advanced Optical Materials",
volume = "8",
number = "20",
pages = "2000786",
year = "2020",
}

@ARTICLE{bib33,
   title = "The third-order nonlinear optical susceptibility of gold",
journal = "Optics Communications",
volume = "326",
pages = "74-79",
year = "2014",
issn = "0030-4018",
author = "R. W. Boyd and Z. Shi and I. De Leon",
}

@phdthesis{bib34,
   author = "Ikenna Onwukwe",
   title = "Nonlinear detection of third harmonic generation (THG) in gold metamaterial ",
   school = "University of Victoria",
   type = "Master thesis",
   address = " Department of Electrical and Computer Engineering",
   month = "September",
   year = "2016",
}

@ARTICLE{bib35,
   author       = "H. Aouani and M. Rahmani and M. Navarro-Cía and S. A. Maier ", 
    title       ="Third-harmonic-upconversion enhancement from a single semiconductor nanoparticle coupled to a plasmonic antenna",
   year         = "2014", 
   journal      = "Nature Nanotechnology", 
   volume       = "9", 
   pages        = "290–294",
}

@ARTICLE{bib36,
   author       = "Y. Pu and R. Grange and C.-L. Hsieh and D. Psaltis", 
    title       ="Nonlinear optical properties of core-shell nanocavities for enhanced second-harmonic generation",
   year         = "2010", 
   journal      = "Phys. Rev. Lett.", 
   volume       = "104", 
   pages        = "207402",
}

@ARTICLE{bib37,
   author       = "H. Aouani and M. Rahmani and M. Navarro-Cía and S. A. Maier ", 
    title       ="Third-harmonic-upconversion enhancement from a single semiconductor nanoparticle coupled to a plasmonic antenna",
   year         = "2014", 
   journal      = "Nature Nanotechnology", 
   volume       = "9", 
   pages        = "290–294",
}

@ARTICLE{bib38,
   author       = "B. Metzger and M. Hentschel and T. Schumacher and M. Lippitz and X. Ye and C. B. Murray and B. Knabe and K. Buse and H. Giessen", 
    title       ="Doubling the efficiency of third harmonic generation by positioning ITO nanocrystals into the hot-spot of plasmonic gap-antennas",
   year         = "2014", 
   journal      = "Nano Letters", 
   volume       = "14", 
   pages        = "2867–2872",
}

@ARTICLE{bib39,
   author       = "T. Shibanuma and G. Grinblat and P. Albella and S. A. Maier", 
    title       ="Efficient third harmonic generation from metal–dielectric hybrid nanoantennas",
   year         = "2017", 
   journal      = "Nano Letters ", 
   volume       = "17", 
   pages        = "2647-2651",
}

@ARTICLE{bib40,
   author       = "M. A. Vincenti and D. de Ceglia and M. Grande and A. D’Orazio and M. Scalora ", 
    title       ="Tailoring Absorption in Metal Gratings with Resonant Ultrathin Bridges",
   year         = "2013", 
   journal      = "Plasmonics", 
   volume       = "8", 
   pages        = "1445–1456",
}

@ARTICLE{bib41,
   author       = "G. D’Aguanno and N. Mattiucci and A. Alù and M. J. Bloemer", 
    title       ="Quenched optical transmission in ultrathin subwavelength plasmonic gratings",
  journal = {Phys. Rev. B},
  volume = {83},
  issue = {3},
  pages = {035426},
  numpages = {11},
  year = {2011},
}

@ARTICLE{bib42,
   author       = "J. Braun and B. Gompf and G. Kobiela and M. Dressel", 
    title       ="How Holes Can Obscure the View: Suppressed Transmission through an Ultrathin Metal Film by a Subwavelength",
  journal = {Phys. Rev. Lett.},
  volume = {103},
  issue = {20},
  pages = {203901},
  numpages = {4},
  year = {2009},
}

@ARTICLE{bib43,
   author       = "I. S. Spevak and A. Yu. Nikitin and E. V. Bezuglyi and Alex Levchenko and A. V. Kats", 
    title       ="Resonantly suppressed transmission and anomalously enhanced light absorption in periodically modulated ultrathin metal films",
   journal = {Phys. Rev. B},
  volume = {79},
  issue = {16},
  pages = {161406},
  numpages = {4},
  year = {2009},
}

\section*{Funding}

SM, JT, and CC acknowledge the Spanish Agencia Estatal de Investigación from Ministerio de Ciencia, Investigación y
Universidades (project PID2023-148620NB-I00 granted by MICIU/AEI/10.13039/501100011033 and FEDER, UE) and US
Army Research Laboratory Cooperative Agreement N◦ W911NF-22-2-0236 issued by US ARMY ACC-APG-RTP. S. M.
acknowledges predoctoral grant FPI-UPC 2022, funded by Universitat Politècnica de Catalunya (Government of Catalunya).
M. A. V. thanks NATO SPS Grant no. G5984 - RESPONDER for financial support. This work has received funding from the
Spanish Agencia Estatal de Investigación/ Spanish Ministry of Science and Innovation (AEI/MCIN) through grants PID2022-
141956NB-I00 MCIN/AEI/10.13039/501100011033 (OUTLIGHT), and CEX2019-000917-S (FUNFUTURE, Spanish Severo
Ochoa Centre of Excellence program) and from the Generalitat de Catalunya (2021-SGR-00444). A. C. R. acknowledges
support from the postdoctoral fellowship program Beatriu de Pinós 2022 BP 00083, funded by the Secretary of Universities and
Research (Government of Catalonia).

\section*{Author contributions statement}

All authors contributed to the sample design. A.C.R. and A.M. fabricated the samples. S.M., C.C. and J.T. conducted the experiment(s) and analyzed the results. S. M. performed the numerical simulation(s) with supervision of all authors. C.C. and A.M. supervised and led the scientific collaboration. All authors contributed to the writing and review of the manuscript. 

\end{document}